\documentclass[aps, pre, reprint, superscriptaddress, amsmath, amssymb]{revtex4-2}

\usepackage{graphicx}% Include figure files
\usepackage{dcolumn}% Align table columns on decimal point
\usepackage{bm}% bold math
\usepackage[utf8]{inputenc}
\usepackage[T1]{fontenc}
\usepackage{hyperref} % keep last; supports \url in refs

\begin{document}

\title{Helical Field-Driven Translational-Rotational Conversion in Conductors}

\author{M. D. Tokman}
\affiliation{Department of Electrical and Electronic Engineering, and Schlesinger Family Center for Compact Accelerators, Radiation Sources and Applications (FEL), Ariel University, 40700 Ariel, Israel}

\author{V. L. Bratman}
\affiliation{Department of Electrical and Electronic Engineering, and Schlesinger Family Center for Compact Accelerators, Radiation Sources and Applications (FEL), Ariel University, 40700 Ariel, Israel}

\author{E. Magori}
\affiliation{Department of Electrical and Electronic Engineering, Jerusalem College of Technology, 91160 Jerusalem, Israel}

\author{N. Balal}
\email{nezahb@ariel.ac.il}
\affiliation{Department of Electrical and Electronic Engineering, and Schlesinger Family Center for Compact Accelerators, Radiation Sources and Applications (FEL), Ariel University, 40700 Ariel, Israel}

\begin{abstract}
A theory of a new magnetodynamic effect describing the energy exchange between the degrees of freedom of a conducting cylinder moving in a helical magnetic field has been developed. The possibility of effectively converting the translational motion of the cylinder into its angular rotation around the system axis (translational-rotational conversion, TRC) has been demonstrated. A connection between this effect and the formation of helical trajectories of electrons in undulators in free electron lasers (FELs) and the inverse Faraday effect (IFE) has been revealed. In TRC, unlike many known effects associated with the interaction of a moving conductor with a magnetic field, the conductor has no electrical contact with any external circuit, which makes it especially attractive for various applications. The TRC is also possible in a magnetohydrodynamic flow moving along the axis of a helical magnetic field. The theory is formulated in the limit of large magnetic Reynolds numbers, which corresponds to a sufficiently fast motion of well-conducting objects. In this scenario, the dynamics of the system is described by a nonlinear pendulum equation or a nonlinear pendulum equation with a nonzero right-hand side. In the latter case, a system dynamic mode corresponding to phase lock can be implemented.
\end{abstract}

\maketitle

\section{Introduction}

The article presents a theory of a new magnetodynamic effect of converting the energy of translational motion of a conducting cylinder moving in a helical magnetic field into the energy of its rotational motion (translational-rotational conversion, TRC). Unlike other effects associated with the interaction of a moving conductor with a magnetic field (for example, such striking examples as a unipolar inductor~\cite{landau1984} or a magnetohydrodynamic generator~\cite{cowling1957}), in TRC the conductor has no electrical contact with any external circuit. The absence of losses typically associated with electrical contacts suggests that this effect has significant prospects for various applications. To some extent, this effect resembles the poloidal rotation of plasma in closed (toroidal) magnetic plasma confinement systems (see, e.g.,~\cite{shaing2003}). Conversely, achieving this effect in magnetic traps critically depends on a system of closed toroidal magnetic surfaces~\cite{krall1973}. Such surfaces are evidently absent in the helical magnetic field considered. A simple analogue of the effect considered in this work is the motion of a free electron in the helical field of a magnetic undulator in free electron lasers (FELs) (see, e.g.,~\cite{freund2024}). In FELs, helical undulators allow the initial translational velocity of free electrons to be converted into rotational velocity. It seems natural to use a similar effect for electrons in a metal and thereby convert the translational velocity of a conductor into rotational velocity. In a conductor, of course, current carriers are intensely scattered by impurities, dislocations, and lattice vibrations. Nevertheless, the ensemble-averaged motion of electrons in such a field still corresponds to the conversion of the translational motion of electrons, caused by the motion of the conductor, into their rotational motion. This effect has a certain analogy with the inverse Faraday effect (IFE)~\cite{landau1984,tokman2020}. The simple models mentioned above are considered in Section~II. Ultimately, the angular momentum of the electron component is transferred to the crystal lattice of the conductor. This process is significantly influenced by the effects of electrodynamics of continuous media. For a moving conductor with high conductivity, the magnetic Reynolds number is relatively large; therefore, the magnetic field within it differs significantly from the field of external sources in a vacuum. This effect determines the formation of current in the conductor and, ultimately, the ponderomotive action of the external field on the conductor. Section~III is devoted to the formulation of the corresponding initial equations in the approximation of ideal magnetohydrodynamics. Finally, in Section~IV, the dynamic equations for the longitudinal motion and rotation of the cylinder about the axis of symmetry are formulated and analyzed. It turns out that they correspond to a system of equations describing the dynamics of a nonlinear oscillator.

A video demonstrating this effect is available as supplementary material~\cite{figshare_video}. A description of the experiment with detailed measurements will be published in a separate article.

\section{Motion of charged particles in a helical magnetic field}

\subsection{Motion of free electrons}

\begin{figure}
\includegraphics[width=\columnwidth]{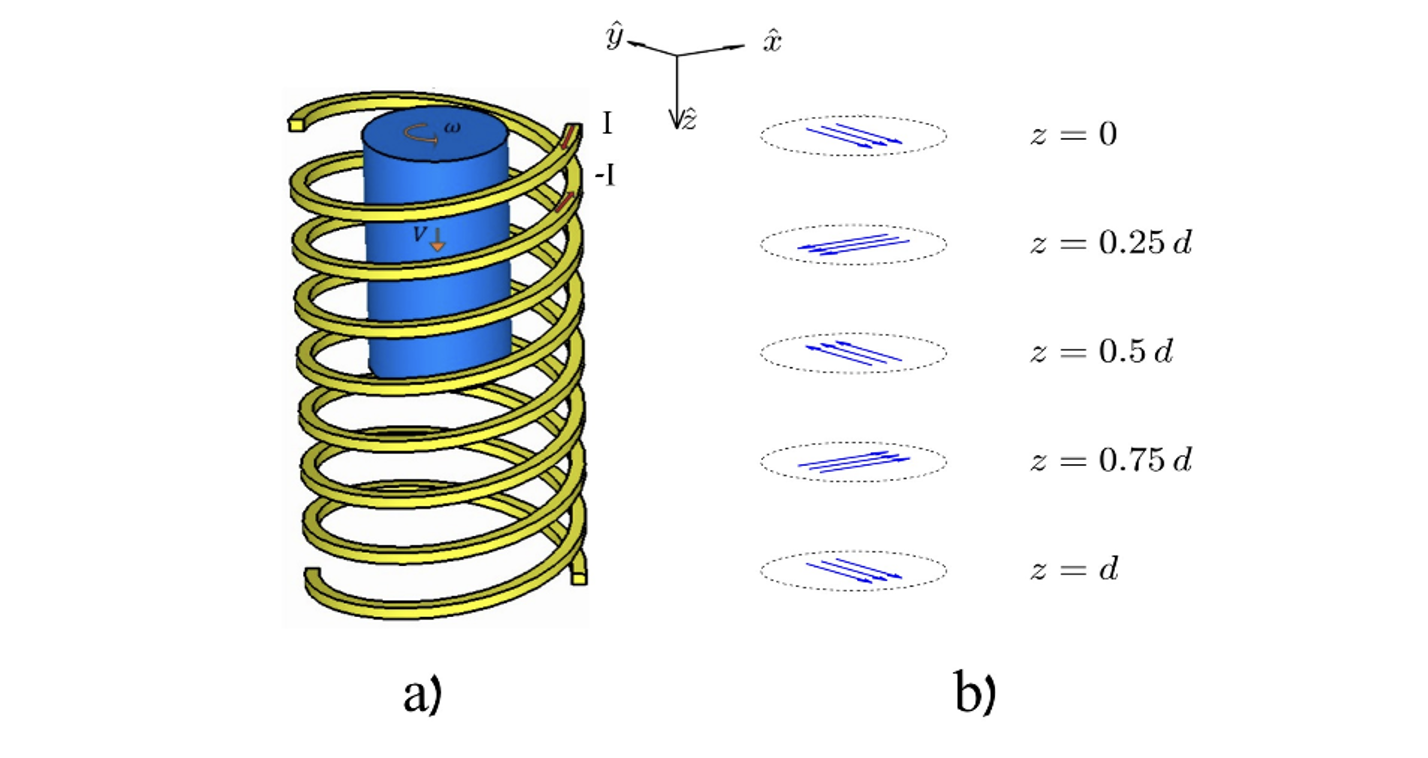}
\caption{\label{fig:setup} a) The conducting cylinder moves along the axis of the helical field (a bifilar system of helical currents is shown as an example of a source of a helical field), b) the helical field described by approximate Eq.~(\ref{eq:helical_field}) is uniform in perpendicular cross sections, being directed perpendicular to the axis; direction $\mathbf{B}$ rotates uniformly when moving along the axis.}
\end{figure}

A helical magnetic field with amplitude $B$ and period $d = 2\pi/h$ (Fig.~\ref{fig:setup}) at a relatively small distance $r$ from the axis (i.e., at $hr \ll 1$) is described in Cartesian coordinates by the simple formula
\begin{equation}
\mathbf{B} = B(\mathbf{x}_0\sin hz - \mathbf{y}_0\cos hz), \label{eq:helical_field}
\end{equation}
where $\mathbf{x}_0$ and $\mathbf{y}_0$ are the unit vectors of the corresponding axes. The nonrelativistic equations of motion of particles with mass $m$ and charge $-e$ in such a field have the form
\begin{align}
\dot{\upsilon}_x &= -\omega_B\upsilon_z\cos hz, \label{eq:motion_x}\\
\dot{\upsilon}_y &= -\omega_B\upsilon_z\sin hz, \label{eq:motion_y}\\
\dot{\upsilon}_z &= \omega_B(\upsilon_x\cos hz + \upsilon_y\sin hz), \label{eq:motion_z}\\
\dot{z} &= \upsilon_z. \label{eq:motion_pos}
\end{align}

Here, $\upsilon_{x,y,z}$ are the components of the particle velocity and $\omega_B = eB/mc$. The forced solution of these equations
\begin{align}
\upsilon_x &= -\upsilon_\bot\sin hz, \label{eq:solution_x}\\
\upsilon_y &= \upsilon_\bot\cos hz, \label{eq:solution_y}\\
\upsilon_z &= \text{const} \label{eq:solution_z}
\end{align}
presents rotation in the transverse plane with the constant undulator frequency $\Omega_u = h\upsilon_z$ and velocity $\upsilon_\bot = \sqrt{\upsilon_x^2 + \upsilon_y^2} = \omega_B/h$ along a circle of radius $\upsilon_\bot/\Omega_u$, and the motion along the $z$ coordinate also occurs at a constant velocity. Such a solution can be implemented, for example, for the initial conditions $\upsilon_{x,y} = 0$ with a sufficiently slow (adiabatically smooth) "switch on" of the magnetic field in space or time.

With an adiabatically smooth change in the magnetic field parameters $B$ and $h$ along the $z$ coordinate, the transverse velocity of the particle also changes adiabatically: $\upsilon_\bot = \frac{\omega_B(z)}{h(z)}$. In this case, the longitudinal velocity $\upsilon_z$ is determined from the law of conservation of energy, $\upsilon_\bot^2 + \upsilon_z^2 = \text{const}$. Thus, when moving in the direction of increasing amplitude $B$ and/or spatial period $d$, the energy of translational motion is converted into the energy of rotation of the particle around an axis parallel to the $z$ axis, whereas in a strictly periodic field these quantities are constant.

\subsection{Motion of a test particle in the case of intense scattering}

When considering the behavior of charged particles under conditions of intense scattering, we use the so-called test particle method~\cite{krall1973,ginzburg1961}. This approach considers the equations of particle motion averaged over the chaotic motion in an ensemble of electrons. These equations include a friction force proportional to the particle velocity and the momentum-transfer frequency $\nu$:
\begin{align}
\dot{\upsilon}_x &= -\omega_B\upsilon_z\cos hz - \nu\upsilon_x, \label{eq:test_x}\\
\dot{\upsilon}_y &= -\omega_B\upsilon_z\sin hz - \nu\upsilon_y, \label{eq:test_y}\\
\dot{\upsilon}_z &= \omega_B(\upsilon_x\cos hz + \upsilon_y\sin hz) - \nu(\upsilon_z - V), \label{eq:test_z}
\end{align}
where $V$ is the velocity of the "ionic core" of the conductor along the $z$ axis. Note that for any initial conditions at times $t \sim \nu^{-1}$ the solution of Eqs.~(\ref{eq:test_x})--(\ref{eq:test_z}) relaxes to the forced regime. For transverse motion, such a solution has the form
\begin{align}
\upsilon_x &= -\upsilon_\bot\cos(hz + \Theta), \label{eq:test_sol_x}\\
\upsilon_y &= -\upsilon_\bot\sin(hz + \Theta), \label{eq:test_sol_y}
\end{align}
where $\upsilon_\bot = \frac{\omega_B\upsilon_z}{\sqrt{\Omega_u^2 + \nu^2}}$, $\cos\Theta = \frac{\nu}{\sqrt{\Omega_u^2 + \nu^2}}$, $\sin\Theta = -\frac{\Omega_u}{\sqrt{\Omega_u^2 + \nu^2}}$. Substituting Eqs.~(\ref{eq:test_sol_x}) and (\ref{eq:test_sol_y}) into Eq.~(\ref{eq:test_z}), we obtain the equation for longitudinal motion
\begin{equation}
\dot{\upsilon}_z = -\frac{\omega_B^2\upsilon_z\cos\Theta}{\sqrt{\Omega_u^2 + \nu^2}} - \nu(\upsilon_z - V) = \nu\left\{V - \left[1 + \frac{\omega_B^2}{\Omega_u^2 + \nu^2}\right]\upsilon_z\right\}. \label{eq:long_motion}
\end{equation}

Equation~(\ref{eq:long_motion}) has the stationary solution
\begin{equation}
\upsilon_z = V\frac{1}{1 + \frac{\omega_B^2}{\Omega_u^2 + \nu^2}}. \label{eq:stationary_sol}
\end{equation}

According to Eq.~(\ref{eq:stationary_sol}), for any value of the magnetic field the longitudinal velocity of the particles is less than the velocity of the ion core of the conductor ($\upsilon_z < V$). Thus, for a finite value of $\nu$ the current carriers slow down the motion of the conductor. In this case, the rotation of the current carriers with frequency $\Omega_u$ described by Eqs.~(\ref{eq:test_sol_x}) and (\ref{eq:test_sol_y}) inevitably leads to the transfer of a torque to the conductor. It is important, however, to note that in a spatially bounded conductor, the role of the charge separation field $\mathbf{E}$ excited in such a system becomes fundamentally important. Obviously, this circumstance should significantly modify the "direct" effect of the formation of rotational motion of current carriers under the action of the Lorentz force.

\subsection{Analogy with the inverse Faraday effect}

From a general physical point of view, the effect under consideration can be considered as a variant of the inverse Faraday effect~\cite{landau1984}, which consists in the generation of a magnetic moment of a substance by a rotating field. Indeed, in the reference frame $K'$, in which the conductor is at rest, along with the rotating magnetic field specified by Eq.~(\ref{eq:helical_field}), there is a rotating electric field, which under the condition $V \ll c$ is given by
\begin{equation}
\mathbf{E}' = \frac{VB}{c}\left(\mathbf{x}_0\cos(hz' + \omega't) + \mathbf{y}_0\sin(hz' + \omega't)\right), \label{eq:electric_field}
\end{equation}
where $\omega' = hV$. According to the theory of the IFE in dissipative media~\cite{tokman2020}, such a field forms in the medium a magnetic moment of unit volume, averaged over the period $\frac{2\pi}{\omega'}$, that is directed along the $z$ axis:
\begin{equation}
\boldsymbol{\mu} = -\mathbf{z}_0\frac{|\sigma(\omega',\nu)|^2|E'|^2}{2en_e\omega'} = -\mathbf{z}_0\frac{|\sigma(\omega',\nu)|^2}{2en_e\omega'}\left(\frac{VB}{c}\right)^2. \label{eq:magnetic_moment}
\end{equation}

Here, $n_e$ is the charge carrier density, $\sigma(\omega',\nu)$ is the complex conductivity, which depends on the field frequency~\cite{landau1984,ginzburg1961}. The above-obtained magnetic moment density determines the angular momentum of the electron component in a unit volume: $\mathbf{n}_\theta = -\frac{m}{e}\boldsymbol{\mu}$. The torque applied to a unit volume of a conductor corresponds to the density of angular momentum transferred to the medium per unit time: $\mathbf{m}_\theta \approx \nu\mathbf{n}_\theta$. In the low-frequency limit, when $\omega' \ll \nu$ and $\sigma \approx \frac{e^2n_e}{m\nu}$, we have
\begin{equation}
\mathbf{m}_\theta \approx \mathbf{z}_0R_m\frac{B^2}{8\pi}, \label{eq:torque_density}
\end{equation}
where $R_m = \frac{4\pi\sigma V}{c^2h}$ is the magnetic Reynolds number~\cite{landau1984} in an unbounded homogeneous medium when the only spatial scale of the system is the period of the helical magnetic field. As mentioned in the Introduction, the approximation of a magnetic field given by external sources corresponds to small Reynolds numbers.

In relation to the IFE in a spatially bounded system it was shown in~\cite{tokman2020} that in the region of sufficiently low frequencies $\omega'$ the influence of the charge separation field on the generation of a magnetic moment is fundamentally important.

\section{Current formation in a conducting cylinder moving in a helical magnetic field}

\subsection{Helical magnetic field in cylindrical coordinates}

In Section~II.1, Eq.~(\ref{eq:helical_field}) for the helical field in a vacuum, which is valid in the dipole approximation $hr \ll 1$, is given in Cartesian coordinates. It is then more convenient to use the cylindrical coordinates $r,\theta,z$ in which Eq.~(\ref{eq:helical_field}) corresponds to
\begin{equation}
\mathbf{B} = B[\mathbf{r}_0\sin(hz - \theta) - \boldsymbol{\theta}_0\cos(hz - \theta)]. \label{eq:cylindrical_field}
\end{equation}

It should be noted that both Eq.~(\ref{eq:helical_field}) and Eq.~(\ref{eq:cylindrical_field}) do not satisfy the equation $\text{rot}\mathbf{B} = 0$ and the transition to the limit $hr \ll 1$ should be done with a certain amount of caution. Let us use the general expression for the helical field:
\begin{equation}
\mathbf{B} = \text{Re}[\mathcal{B}(r)e^{ihz - i\theta}], \label{eq:general_helical}
\end{equation}
where the vector function $\mathcal{B}(r)$ satisfies the condition: $\text{div}(\mathcal{B}(r)e^{ihz - i\theta}) = 0$. In the special case of a field formed by external sources in a vacuum, we have:
\begin{equation}
\mathcal{B}(r) = 2B[-\mathbf{r}_0iI_1'(hr) - \boldsymbol{\theta}_0\frac{I_1(hr)}{hr} + \mathbf{z}_0I_1(hr)], \label{eq:bessel_field}
\end{equation}
where $I_1(hr)$ and $I_1'(hr)$ is a modified Bessel function of the first kind and its derivative.

\subsection{Magnetic Reynolds number}

Let a conducting cylinder move along the $z$ axis with velocity $V$ (Fig.~\ref{fig:setup}a). Assume that the height $\mathcal{L}$ of the cylinder significantly exceeds both its radius $R$ and the spatial period $d$ of the helical magnetic field. In this case, the current formation at the ends of the cylinder should not considerably affect its motion. Assume that the cylinder can also rotate around the $z$ axis with frequency $\omega$. Then for the Euler velocity field of the material points that make up the cylinder we have
\begin{equation}
\boldsymbol{\upsilon} = \mathbf{z}_0V + \boldsymbol{\theta}_0\omega r. \label{eq:velocity_field}
\end{equation}

The magnetic field in a moving conductor can be described by the equation of magnetohydrodynamics~\cite{landau1984}
\begin{equation}
\frac{\partial}{\partial t}\mathbf{B} = \text{rot}(\boldsymbol{\upsilon} \times \mathbf{B}) + \frac{c^2}{4\pi\sigma}\Delta\mathbf{B}. \label{eq:mhd}
\end{equation}

The solution to this equation depends on the magnetic Reynolds number
\begin{equation}
\mathcal{R}_m \sim \frac{|\text{rot}(\boldsymbol{\upsilon} \times \mathbf{B})|}{|\frac{c^2}{4\pi\sigma}\Delta\mathbf{B}|} \sim \frac{4\pi\sigma\upsilon L}{c^2},
\end{equation}
where $\upsilon$ and $L$ are characteristic velocity and scale of the inhomogeneity, respectively. In the case under consideration, there are two scales of system inhomogeneity, namely, $R$ and $h^{-1}$, and two characteristic velocities, $V$ and $\omega R$. Therefore, it turns out to be important to estimate the Reynolds parameter considering the helical symmetry of the magnetic field and the velocity field given by Eq.~(\ref{eq:velocity_field}). Using Eqs.~(\ref{eq:general_helical}) and (\ref{eq:velocity_field}), after straightforward calculations, the following expression can be obtained:
\begin{equation}
\text{rot}(\boldsymbol{\upsilon} \times \mathbf{B}) = -\text{Re}[\mathcal{B}(r)(V\frac{\partial}{\partial z} + \omega\frac{\partial}{\partial\theta})e^{ihz - i\theta}], \label{eq:curl_v_cross_B}
\end{equation}
from which it follows
\begin{equation}
|\text{rot}(\boldsymbol{\upsilon} \times \mathbf{B})| = |\mathbf{B}||hV - \omega|. \label{eq:curl_magnitude}
\end{equation}

As for the second term on the right-hand side of Eq.~(\ref{eq:mhd}), we estimate it by setting $hR < 1$ or $hR \sim 1$:
\begin{equation}
\left|\frac{c^2}{4\pi\sigma}\Delta\mathbf{B}\right| \sim \frac{c^2|\mathbf{B}|}{4\pi\sigma R^2}.
\end{equation}

As a result, we have
\begin{equation}
\mathcal{R}_m = \frac{4\pi\sigma R^2|hV - \omega|}{c^2}. \label{eq:reynolds_number}
\end{equation}

The Eq.~(\ref{eq:reynolds_number}) can be interpreted within the framework of the skin layer theory. Indeed, the standard expression for the skin layer depth~\cite{landau1984} has the form $\delta \approx \sqrt{\frac{c^2}{4\pi\tilde{\omega}}}$, where $\tilde{\omega}$ is the frequency of magnetic field oscillations. For the system under consideration, the frequency of oscillations of the magnetic field phase in the reference frame in which the conductor is at rest should be chosen as $\tilde{\omega} = \left|\frac{d(hz - \theta)}{dt}\right| = |hV - \omega|$. Then the magnetic Reynolds number is equal to the square of the ratio of the cylinder radius to the skin layer size. according to which the Reynolds number is the ratio of the characteristic decay time of Foucault currents $\tau_d \sim \frac{4\pi\sigma R^2}{c^2}$~\cite{landau1984} to the characteristic time of their excitation $\tau_e \sim \tilde{\omega}^{-1}$.\footnote{Estimates for the characteristic skin layer depth $\delta$ and decay time of Foucault currents $\tau_d$ are given for the conditions $hR < 1$ or $hR \sim 1$.}

Thus, for $\mathcal{R}_m \ll 1$, the magnetic field in the conductor corresponds to the field of external sources in a vacuum, and for $\mathcal{R}_m \gg 1$, Foucault currents screen the bulk region of the conductor from the external magnetic field. However, this does not mean that for large magnetic Reynolds numbers, the magnetic field in the conductor must necessarily be zero. Indeed, in the limit $\mathcal{R}_m \gg 1$ Eq.~(\ref{eq:mhd}) yields an equation describing "freezing" of the magnetic field into a moving conductor~\cite{landau1984}:
\begin{equation}
\frac{\partial}{\partial t}\mathbf{B} = \text{rot}(\boldsymbol{\upsilon} \times \mathbf{B}). \label{eq:frozen_field}
\end{equation}

Within the "freezing" concept, the magnetic field in a bulk region at any moment in time is determined by the field that existed in the conductor at the initial moment. Let, for example, the conductor be at rest at first, i.e., $V = \omega = \mathcal{R}_m = 0$, and an external magnetic field penetrates it. If after this a sufficiently fast motion begins, then during this motion the initial magnetic field is "frozen" into the conductor, i.e., it shifts and rotates together with the shift and rotation of the sample. Indeed, for the initial conditions $\mathbf{B}(t = 0) = \text{Re}[\mathcal{B}(r)\exp(ihz - i\theta)]$, using Eq~(\ref{eq:curl_v_cross_B}), we obtain the following solution of equation~(\ref{eq:frozen_field}):
\begin{equation}
\mathbf{B} = \text{Re}\left\{\mathcal{B}\exp\left[ih\left(z - \int_0^t V dt\right) - i\left(\theta - \int_0^t \omega dt\right)\right]\right\} \label{eq:frozen_solution}
\end{equation}
where the quantities $V$ and $\omega$ generally depend on time. If the conductor begins a fast motion instantly, so that the magnetic field does not have time to penetrate it, then the magnetic field in the conductor is indeed equal to zero.

Using an estimate $|hV - \omega| \sim hV$, we obtain a convenient expression for numerical evaluation:
\begin{equation}
\mathcal{R}_m \approx 0.37 \cdot \left(\frac{\sigma}{\sigma_{Cu}}\right)(hR)V(\text{m/s})R,
\end{equation}
where $\sigma_{Cu}$ is the conductivity of copper. With magnetic field period $d \sim R$ for a well-conducting cylinder with a characteristic size of 1 cm, a fairly large value of $\mathcal{R}_m \approx 4.6$ corresponds to a velocity of about 1 m/s (such a velocity is achieved due to the gravitational acceleration when moving away to a distance of 5 cm).

\subsection{Formation of the current system in the "frozen-in" magnetic field regime}

The field inside the conductor is a superposition of the field $\mathbf{B}$ of external sources in a vacuum and the field $\mathbf{B}_{in}$ formed by currents $\mathbf{J}$ flowing inside the conductor:
\begin{equation}
\text{rot}(\mathbf{B}_{in} - \mathbf{B}) = \frac{4\pi}{c}\mathbf{J}. \label{eq:current_relation}
\end{equation}

Although we consider a non-magnetic medium, it is convenient to specify the current using the effective magnetization $\mathbf{M}$ of the cylinder, which we define as follows:
\begin{equation}
4\pi\mathbf{M} = (\mathbf{B}_{in} - \mathbf{B})1(R - r), \label{eq:magnetization}
\end{equation}
where $\mathbf{J} = c\text{rot}\mathbf{M}$ and $1(x)$ is a unit Heaviside function.

As noted above, within the "frozen-in" concept, it is easy to find the current and magnetic field inside the sample if the magnetic field in the sample at the initial moment $\mathbf{B}_0(r,\theta,z)$ is known. In the "frozen-in" regime, when the cylinder is displaced along the $z$ axis by the length $l = \int_0^t V dt$ and rotated by the angle $\varphi = \int_0^t \omega dt$ around this axis, in accordance with Eq.~(\ref{eq:frozen_solution}) we have
\begin{equation}
\mathbf{B}_{in}(r,\theta,z) = \mathbf{B}_0(r,\theta - \varphi,z - l). \label{eq:internal_field}
\end{equation}

Under the condition $\text{rot}\mathbf{B}_0(r < R) = 0$, we obtain $\text{rot}\mathbf{B}_{in}(r < R) = 0$, i.e., in this case there is only a surface current $\mathbf{J}_S$ at the boundary of the conductor. For the effective magnetization $\mathbf{M}$ of the cylinder and the surface current $\mathbf{J}_S$, we obtain
\begin{align}
\mathbf{M} &= \frac{1}{4\pi}(\mathbf{B}_0(r,\theta - \varphi,z - l) - \mathbf{B}(r,\theta,z)), \label{eq:magnetization_explicit}\\
\mathbf{J}_S &= \boldsymbol{\theta}_0M_z(r = R,\theta,z) - \mathbf{z}_0M_\theta(r = R,\theta,z). \label{eq:surface_current}
\end{align}

Here, the function $\mathbf{J}_S(r = R,\theta,z)$ determines the distribution of Foucault currents on the surface of the conductor.

Note that the current $\mathbf{J}_S$ forms a field not only inside, but also outside the conductor. i.e., the magnetic field in the region $r > R$ differs from that of the field of external sources in a vacuum given by Eqs.~(\ref{eq:general_helical}) and (\ref{eq:bessel_field}). The latter circumstance, however, will not affect the motion of the rigid conductor as a whole, although it leads to the occurrence of internal stresses in external sources of the magnetic field.

\section{Motion of a conducting cylinder in a helical magnetic field}

\subsection{Ponderomotive action of a magnetic field on a cylinder for large Reynolds numbers}

The motion of the conductor is determined by the Ampere force acting on the surface current $\mathbf{J}_S$, specified by Eq.~(\ref{eq:surface_current}), or (which is completely equivalent~\cite{landau1984,tamm1979}) through the effective magnetization $\mathbf{M}$, given by Eq.~(\ref{eq:magnetization_explicit}). Within this approach, the force in the direction of the $z$-axis and the moment of forces relative to this axis for a cylindrical element with a unit height $\Delta L = 1$ are determined by the expressions~\cite{tamm1979}
\begin{align}
\mathcal{F}_\parallel &= \iint d^2r\left(\mathbf{M}\frac{\partial}{\partial z}\mathbf{B}\right), \label{eq:force_parallel}\\
\mathcal{M}_\theta &= \iint d^2r\left(\mathbf{M}\frac{\partial}{\partial\theta}\mathbf{B}\right), \label{eq:moment_theta}
\end{align}
where the integration is performed over the cross section of the conductor and the external field $\mathbf{B}(r,\theta,z)$ is given by Eqs.~(\ref{eq:general_helical}) and (\ref{eq:bessel_field}). Following Section~III.2, we consider two cases.

\textbf{(i)} Initially, the conductor was at rest and the external field penetrated the conductor before the start of the rapid motion stage. In this case, for the vector $\mathbf{M}$ in Eq.~(\ref{eq:magnetization_explicit}) we have
\begin{align}
\mathbf{B}_0(r,\theta,z) &= \mathbf{B}(r,\theta,z),\\
\mathbf{B}_0(r,\theta - \varphi,z - l) &= \text{Re}[\mathcal{B}(r)e^{i(h - l)z - i(\theta - \varphi)}].
\end{align}

From these formulas it follows
\begin{align}
\mathcal{F}_\parallel &= -\frac{hb^2R^2}{4}\sin(hl - \varphi), \label{eq:force_result}\\
\mathcal{M}_\theta &= \frac{b^2R^2}{4}\sin(hl - \varphi), \label{eq:moment_result}
\end{align}
where $b^2 = 2R^{-2}\int_0^R |\mathcal{B}(r)|^2 r dr$. In the limit $hR \ll 1$, when $-i\mathcal{B}_r = \mathcal{B}_\theta = B$, $\mathcal{B}_z = 0$, we have $b = B$. In this case, the magnitude of the torque applied to a unit volume is of the same order of magnitude as the magnetic field pressure: $\frac{\sqrt{\langle\mathcal{M}_\theta^2\rangle}}{\pi R^2} \approx \frac{B^2}{8\pi}$ (here, the angle brackets denote averaging over the phase $hl - \varphi$).

\textbf{(ii)} The conductor starts moving instantly, so that the magnetic field does not have time to penetrate it. In this case, $\mathbf{B}_0 = 0$ and
\begin{equation}
\mathcal{F}_\parallel = \mathcal{M}_\theta = 0. \label{eq:instant_motion}
\end{equation}

It is important that in the case of non-uniform amplitude of the helical field (for example, when the sample enters the region of a localized field) $\mathcal{F}_\parallel,\mathcal{M}_\theta \neq 0$ at $\mathbf{B}_0 = 0$. In particular, this option should be considered if the length of the magnetic system is less than the height of the cylinder. However, we limit ourselves here to the case of a purely periodic field.

\subsection{Equations of motion}

From the expressions for the force and moment of forces, given by Eqs.~(\ref{eq:force_result}) and (\ref{eq:moment_result}), we obtain the following equations of motion of a cylinder with a uniform density $\rho$:
\begin{align}
\dot{l} &= V, \label{eq:motion_l}\\
\dot{\varphi} &= \omega, \label{eq:motion_phi}\\
\dot{\omega} &= \aleph^2\sin(hl - \varphi), \label{eq:motion_omega}\\
\dot{V} &= -\frac{hR^2}{2}\aleph^2\sin(hl - \varphi), \label{eq:motion_V}
\end{align}
where $\aleph^2 = \frac{b^2}{2\pi\rho R^2}$. The following conservation law corresponds to Eqs.~(\ref{eq:motion_l})--(\ref{eq:motion_V}):
\begin{equation}
I\frac{\omega^2}{2} + m\frac{V^2}{2} - \mathcal{V}\frac{b^2}{4\pi}\cos(hl - \varphi) = \text{const},
\end{equation}
where $-\mathcal{V}\frac{b^2}{4\pi}\cos(hl - \varphi)$ is the interaction energy of the external magnetic field with the currents in the conductor, $\mathcal{V}$, $m$, and $I$ are the volume, mass, and moment of inertia of the cylinder, respectively. Equations~(\ref{eq:motion_omega}) and (\ref{eq:motion_V}) give the relationship between the changes in longitudinal and angular velocities: $h(V - V_0) = -\epsilon(\omega - \omega_0)$, where $\epsilon = \frac{h^2R^2}{2}$. Using Eqs.~(\ref{eq:motion_l})--(\ref{eq:motion_V}), we obtain the equation of a nonlinear pendulum for the phase $\chi = hl - \varphi$:
\begin{equation}
\ddot{\chi} + \aleph^2(1 + \epsilon)\sin\chi = 0. \label{eq:pendulum}
\end{equation}

Moving on to dimensionless variables
\begin{align}
\tau &= \aleph\sqrt{1 + \epsilon} t, \label{eq:dim_tau}\\
\zeta &= \frac{\omega}{\aleph\sqrt{1 + \epsilon}}, \label{eq:dim_zeta}\\
\nu &= \frac{hV}{\aleph\sqrt{1 + \epsilon}}, \label{eq:dim_nu}
\end{align}
we arrive at the system of equations
\begin{align}
\frac{d^2\chi}{d\tau^2} + \sin\chi &= 0, \label{eq:pendulum_dim}\\
\frac{d\chi}{d\tau} &= \nu - \zeta, \label{eq:phase_velocity}\\
\nu - \nu_0 &= -\epsilon(\zeta - \zeta_0), \label{eq:velocity_relation}
\end{align}
having the integral
\begin{equation}
(\nu - \zeta)^2 - 2\cos\chi = \text{const}. \label{eq:integral}
\end{equation}

\begin{figure}
\includegraphics[width=\columnwidth]{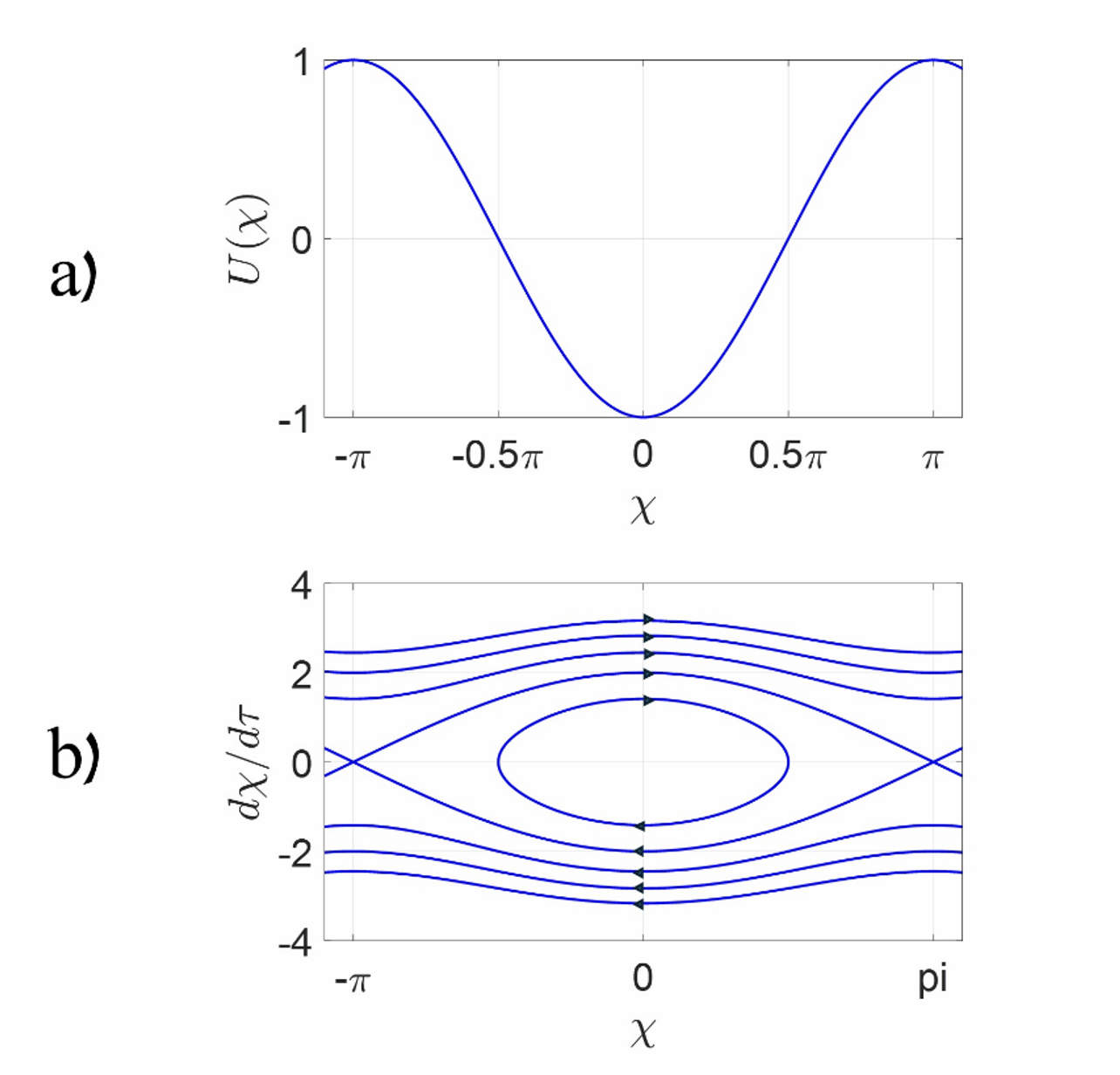}
\caption{\label{fig:phase_plane} a) Potential energy $U(\chi) = -\cos\chi$ and b) the phase plane of a nonlinear pendulum.}
\end{figure}

On the phase plane corresponding to Eq.~(\ref{eq:pendulum_dim}), there are two equilibrium states, stable $\dot{\chi} = \chi = 0$ and unstable $\dot{\chi} = 0$, $\chi = \pi$ (Fig.~\ref{fig:phase_plane}).

For initial conditions with zero rotation frequency
\begin{equation}
\chi_0 = 0, \quad \zeta_0 = 0, \quad \nu_0 \neq 0
\end{equation}
from Eqs.~(\ref{eq:velocity_relation}) and (\ref{eq:integral}) it follows
\begin{align}
\zeta &= \frac{1}{1 + \epsilon}\left(\nu_0 \mp \sqrt{\nu_0^2 - 2(1 - \cos\chi)}\right), \label{eq:zeta_solution}\\
\nu &= \frac{1}{1 + \epsilon}\left(\nu_0 \pm \epsilon\sqrt{\nu_0^2 - 2(1 - \cos\chi)}\right). \label{eq:nu_solution}
\end{align}

The upper or lower signs in these equations must be chosen simultaneously. Under the condition $\nu_0 < 2$ the phase $\chi$ periodically changes in the range from $\arccos\left(1 - \frac{\nu_0^2}{2}\right)$ to $-\arccos\left(1 - \frac{\nu_0^2}{2}\right)$ This is the so-called closed phase trajectory regime (Fig.~\ref{fig:phase_plane}). For such trajectories, the normalized longitudinal and angular velocities oscillate in intervals from $\nu_0$ to $\nu_0\frac{1 - \epsilon}{1 + \epsilon}$ and from 0 to $\frac{2\nu_0}{1 + \epsilon}$, respectively. In the parameter range $\nu_0 > 2$, the phase $\chi$ changes monotonically, and the corresponding intervals of change in dimensionless longitudinal velocity and frequency are as follows: from $\nu_0$ to $\frac{\nu_0 + \epsilon\sqrt{\nu_0^2 - 4}}{1 + \epsilon}$ and from 0 to $\frac{\nu_0 - \sqrt{\nu_0^2 - 4}}{1 + \epsilon}$, respectively. Under the condition $\epsilon \ll 1$, the "spinning" of the cylinder does not lead to a significant loss of longitudinal velocity.

Let us consider as an example the efficiency of magnetodynamic uncoiling of a copper cylinder ($\rho \approx 9$ g/cm$^3$) with a radius $R = 1$ cm, moving with an initial velocity $V_0 = 5$ m/s in a helical magnetic field with a period $d = 5$ cm ($\epsilon \approx 0.8$) and a field amplitude $B_0 = 1.2$ kG on the system axis. Numerical evaluation using expressions~(\ref{eq:zeta_solution}), (\ref{eq:nu_solution}) leads to the following result: during time $t \approx 30$ ms the longitudinal velocity of the cylinder decreases from 5 to 1 m/s, and the cylinder spins up to $\omega \approx 400$ s$^{-1}$. In this case, 99\% of the translational kinetic energy is converted into rotational energy.

In conclusion of this section, we make an important remark related to the applicability of the model used. It should be noted that the expression for the magnetic Reynolds number, given by Eq.~(\ref{eq:reynolds_number}), can be represented as $\mathcal{R}_m = \frac{4\pi\sigma R^2|\dot{\chi}|}{c^2}$. From the latter expression, it follows that for closed phase trajectories the value $\mathcal{R}_m = \dot{\chi} = 0$ is inevitably reached. In this case, formally, the condition of "freezing-in" of the magnetic field is not met. However, since the initial value of the magnetic Reynolds number $\mathcal{R}_{m0} = \frac{4\pi\sigma R^2hV_0}{c^2}$ is known to be large, the relatively short stay of the system in the region of small magnetic Reynolds numbers cannot significantly affect the magnetic field "frozen" into the conductor. Indeed, the change in the magnetic field is determined by the ratio between the characteristic decay time $\tau_d$ of the magnetic field in the conductor and the characteristic time the system spends in the region where $|\dot{\chi}| \ll |\dot{\chi}(t = 0)| = hV_0$. The last interval, in turn, is certainly less than the characteristic oscillation time of "velocity" $\dot{\chi}$, which can be estimated as $T \sim \frac{1}{\aleph\sqrt{1 + \epsilon}}$. The quantity $\tau_d$ corresponds to the decay time of eddy currents, the expression for which is given in Section~III.1. Taking into account the condition for the realization of closed trajectories, we obtain $\frac{T}{\tau_d} \sim \frac{1}{\mathcal{R}_{m0}} \ll 1$. Of course, all the above does not apply to the phase trajectories of the system passing in a small neighborhood of the separatrix (i.e., a trajectory passing through an unstable equilibrium state (Fig.~\ref{fig:phase_plane}), since in this case the time spent near the point $\dot{\chi} = 0$, $\chi = \pi$ tends to infinity. However, even for such trajectories (which correspond to a relatively small proportion of possible initial conditions), the simple model proposed in this work is applicable at times $t < \tau_d$, which for good conductors is quite sufficient to describe the regime of significant "winding" of the cylinder to a frequency $\omega \approx hV$.

\subsection{Motion in the presence of an external force}

Consider a scenario where, in addition to the Ampere force, an external force $F$ acts on the cylinder along the axis of the system. Using the dimensionless variables introduced in the previous section, we obtain the following system of equations:
\begin{equation}
\frac{d^2\chi}{d\tau^2} + \sin\chi - \alpha = 0, \label{eq:pendulum_force}
\end{equation}
where $\alpha = \frac{hF}{M\aleph^2(1 + \epsilon)}$, where $M$ is the mass of the cylinder,
\begin{align}
\frac{d\zeta}{d\tau} &= \frac{1}{1 + \epsilon}\sin\chi, \label{eq:zeta_force}\\
\frac{d\nu}{d\tau} &= -\frac{\epsilon}{1 + \epsilon}\sin\chi + \alpha. \label{eq:nu_force}
\end{align}

A potential profile corresponding to Eq.~(\ref{eq:pendulum_force}) is $U(\chi) = -(\cos\chi + \alpha\chi)$ (Fig.~\ref{eq:nu_force}). Considering Eq.~(\ref{eq:phase_velocity}), we obtain the following integral of motion:
\begin{equation}
(\nu - \zeta)^2 - 2\cos\chi - 2\alpha\chi = \text{const} \label{eq:integral_force}
\end{equation}

In the general case, the separation into regimes with periodic and aperiodic changes in phase $\chi$ depends on the initial conditions (Fig.~\ref{eq:nu_force}). However, periodic motion is impossible in any case where $\alpha > 1$. Let us discuss, for example, the initial conditions: $\chi_0 = \zeta_0 = \nu_0 = 0$. In this case, periodic motion is possible under the condition
\begin{equation}
1 + \sqrt{1 - \alpha^2} > \alpha(\pi - \arcsin(\alpha)), \label{eq:periodic_condition}
\end{equation}
i.e. for $\alpha < \alpha_{cr}$, where $\alpha_{cr} \approx 0.72$. It is easy to see that in the case of periodic motion under specified initial conditions, the value of $\sin\chi(\tau)$ is always positive.

\begin{figure}
\includegraphics[width=\columnwidth]{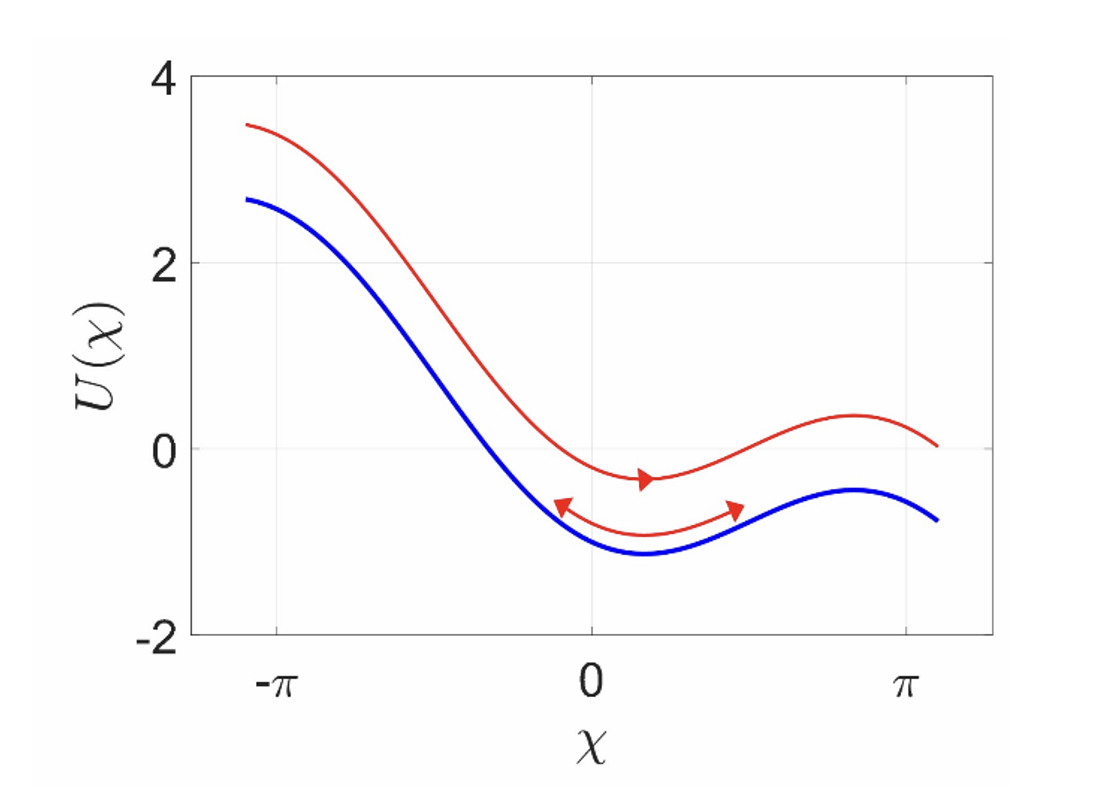}
\caption{\label{fig:potential_force} Potential energy $U(\chi)$ for a nonlinear pendulum with a pulling field ($\alpha=0.5$). Periodic and aperiodic phase oscillations are shown.}
\end{figure}

\begin{figure}
\includegraphics[width=\columnwidth]{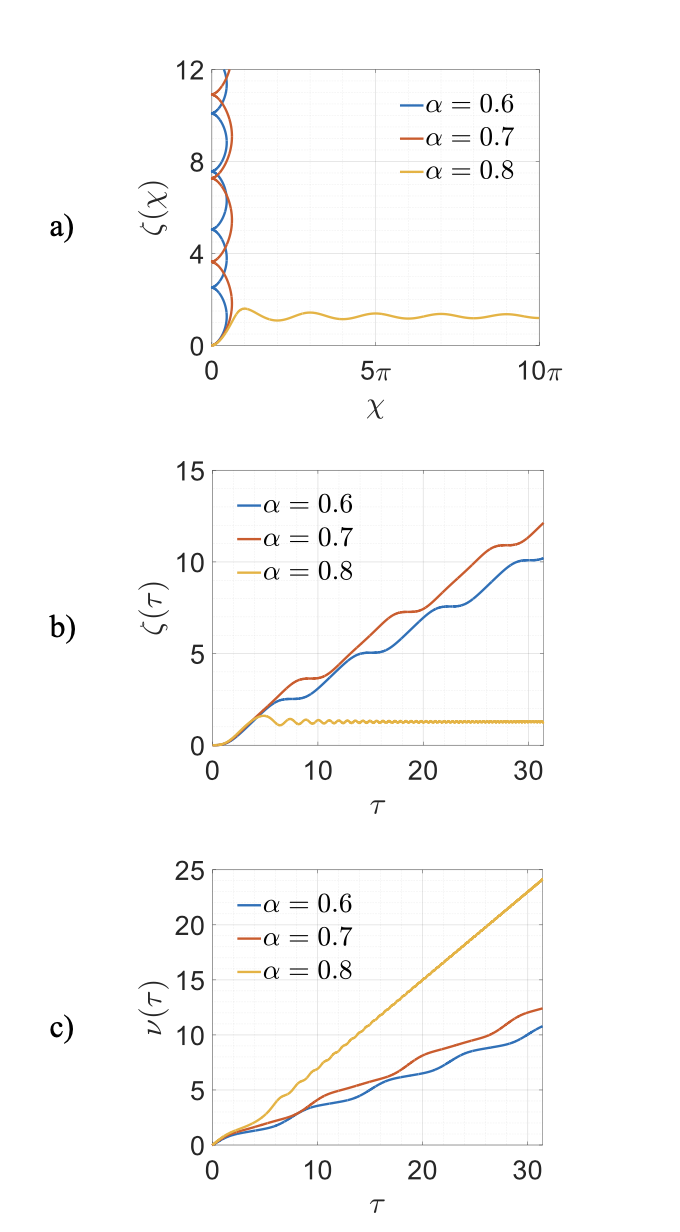}
\caption{\label{fig:calculations} Results of calculations in dimensionless variables: a) and b) angular velocity as a function of phase and time, respectively, and c) translational velocity as a function of time.}
\end{figure}

Then from Eq.~(\ref{eq:zeta_force}), it follows that the angular velocity of the system grows unboundedly. Such a mode of motion corresponds to the phase-locking effect~\cite{deutsch1991,friedland1992,litvak1993,lindberg2004,polomarov2007,tokman2011}. At the same time, in the regime with aperiodic change in phase $\chi$, the angular velocity tends to some limiting value. The results of the corresponding numerical calculations are shown in Fig.~\ref{fig:calculations}.

\section{Conclusions}

In this paper, we theoretically study a new magnetodynamic effect that appears in the coupling between the translational and rotational degrees of freedom of a cylindrical conducting sample. The described translational-rotational conversion (TRC) effect occurs when the cylinder moves in a helical magnetic field that coincides with the field of a helical undulator in free-electron lasers (FELs). A similar effect is also expected in a magnetohydrodynamic flow moving along the axis of a helical magnetic field.

The developed theory is based on approximations of a purely periodic helical field and ideal magnetohydrodynamics at high magnetic Reynolds numbers. Naturally, specific experiments may require going beyond these limitations and further development of the theory. We consider it necessary to discuss here the qualitative aspects of the effect at low Reynolds numbers. As the conductor velocity decreases, leading to a reduced magnetic Reynolds number, the external magnetic field penetrates the bulk conductor region. When $\mathcal{R}_m \ll 1$, the magnetic field can be considered as coinciding with that produced by external sources in a vacuum. It can be demonstrated that, given the evident smallness of kinematic parameters ($hV$ and $\omega$) compared to the metal conductivity ($\sigma$), the ponderomotive force and the corresponding torque at low Reynolds numbers are proportional to the small parameter $\mathcal{R}_m$. For a straightforward comparison, refer to Eqs.~(\ref{eq:torque_density}) and (\ref{eq:moment_result}) and related comments (a more detailed discussion of ponderomotive effects in the low Reynolds number regime, as well as experimental validation, will be presented in a separate publication).

Contactless energy exchange between the translational and rotational degrees of freedom of a macroscopic object within a magnetic field is an intriguing physical phenomenon. The helical field geometry naturally couples these degrees of freedom, enabling momentum injection without the need for sliding electrical contacts or mechanical gears. We believe that this effect has significant potential for various applications~\cite{balal2025}.

\begin{acknowledgments}
This research was supported by the Ministry of Innovation, Science \& Technology, Israel.

We thank Moshe Klein for his valuable technical assistance in developing the experimental demonstration system.
\end{acknowledgments}

\end{document}